\documentclass[runningheads]{llncs}
\usepackage{graphicx,amsmath,multirow,bm,color,xr}
\usepackage{graphicx,amstext,amsfonts,bm}
\usepackage[misc]{ifsym}
\usepackage[colorlinks,urlcolor=blue, citecolor=blue, linkcolor=blue]{hyperref}
\makeatletter \def\thanks#1{\protected@xdef\@thanks{\@thanks \protect\footnotetext{#1}}} \makeatother
\begin{document}

\title{Learning towards Synchronous Network Memorizability and Generalizability for Continual Segmentation across Multiple Sites}

\titlerunning{SMG-Learning}

\author{Jingyang Zhang\inst{1}\thanks{J. Zhang and P. Xue - Equal contribution.}\and
Peng Xue\inst{1} \and
Ran Gu\inst{2} \and
Yuning Gu\inst{1} \and
Mianxin Liu\inst{1} \and
Yongsheng Pan\inst{1} \and
Zhiming Cui\inst{1,3} \and
Jiawei Huang\inst{1} \and
Lei Ma\inst{1} \and
Dinggang Shen \inst{1,4}$^{(\textrm{\Letter})}$}

\authorrunning{J. Zhang et al.}

\institute{School of Biomedical Engineering, ShanghaiTech University, Shanghai, China
\email{dgshen@shanghaitech.edu.cn} \\
\and
School of Mechanical and Electrical Engineering, University of Electronic Science and Technology of China, Chengdu, China \and
Department of Computer Science, The University of Hong Kong, Hong Kong, China \and
Shanghai United Imaging Intelligence Co., Ltd., Shanghai, China}

\maketitle

\begin{abstract}
In clinical practice, a segmentation network is often required to continually learn on a sequential data stream from multiple sites rather than a consolidated set, due to the storage cost and privacy restriction.
However, during the continual learning process, existing methods are usually restricted in either network memorizability on previous sites or generalizability on unseen sites.
This paper aims to tackle the challenging problem of Synchronous Memorizability and Generalizability (SMG) and to simultaneously improve performance on both previous and unseen sites, with a novel proposed SMG-learning framework.
First, we propose a Synchronous Gradient Alignment (SGA) objective, which \emph{not only} promotes the network memorizability by enforcing coordinated optimization for a small exemplar set from previous sites (called replay buffer), \emph{but also} enhances the generalizability by facilitating site-invariance under simulated domain shift.
Second, to simplify the optimization of SGA objective, we design a Dual-Meta algorithm that approximates the SGA objective as dual meta-objectives for optimization without expensive computation overhead.
Third, for efficient rehearsal, we configure the replay buffer comprehensively considering additional inter-site diversity to reduce redundancy.
Experiments on prostate MRI data sequentially acquired from six institutes demonstrate that our method can simultaneously achieve higher memorizability and generalizability over state-of-the-art methods.
Code is available at \url{https://github.com/jingyzhang/SMG-Learning}.

\end{abstract}
\section{Introduction}
Data aggregation of multiple clinical sites \cite{9000851,dhruva2020aggregating} is desired to train convolutional neural networks for medical image segmentation \cite{zhang2020weakly,nie20183,xiang2018deep}.
Compared with aggregating multi-site datasets as a large-scaled consolidated set, training network on a sequential data stream is more storage-efficient and privacy-preserving \cite{zhang2021comprehensive}, where data of different sites arrives in sequence without consuming storage of most (or even all) data of previous sites \cite{li2020continual}.
However, consecutively fine-tuning model on only the data from different incoming sites would confront two challenges:
1) weak memorizability \cite{forgetting}, causing significant performance drop on previously learned sites with data discrepancy due to their different imaging protocols;
2) poor generalizability \cite{liu2021feddg}, decreasing performance on unseen sites with out-of-distribution data \cite{koh2021wilds} and thus impeding direct model deployment in clinical practice \cite{chen2018semantic}.
Therefore, a \emph{Synchronous Memorizability and Generalizability (SMG)} is desirable for the network to simultaneously improve performance on both previous and unseen sites during consecutive training on a data stream.

However, prevalent researches have identified network memorizability and generalizability as two isolated tasks with customized solutions, respectively.
Specifically, Continual Learning (CL) \cite{delange2021continual} is proposed to enhance memorizability and mitigate forgetting for old knowledge of previously learned sites \cite{zhang2021comprehensive}.
CL usually relies on the storage of a small exemplar set from previous sites (called replay buffer) for rehearsal \cite{rebuffi2017icarl,li2020continual,castro2018end}. Then a similarity constraint \cite{lopez2017gradient,gupta2020maml} is imposed between the gradient directions of losses on the replay buffer and the incoming site, enabling coordinated optimization for previous sites and thus relieving the model forgetting.
Besides, Domain Generalization (DG) \cite{zhou2021domain} aims to make a model generalize well to unseen sites, with a key aspect to consider learning invariance \cite{ganin2016domain} across multiple sites with data discrepancy.
In DG, the replay buffer collaborated with the data of incoming site serves as a union dataset to provide multi-site distribution for model training.
Notably, it is arbitrarily split into virtual-train and virtual-test subsets within each minibatch to simulate real-world domain shift \cite{li2018learning,dou2019domain}, boosting feature invariance for network generalizability.

Although above-mentioned methods exhibit advantages on either network memorizability or generalizability, none of them is capable of \emph{Learning with SMG (i.e., SMG-Learning)} that gains both advantages simultaneously \cite{riemer2018learning} due to the following three reasons.
First, for network memorizability, CL methods determine coordinated gradient for previous sites by optimizing a unidirectional transfer to the incoming site, which yet biases the learning process and hampers generalization capacity.
Second, DG methods enhance feature invariance for a generalized model by using arbitrary data splitting to construct cross-site discrepancy, while without a particular consideration of performance balance between previous and incoming sites to consistently maintain network memorizability.
Third, the replay buffer, used in both methods, is commonly populated with representative exemplars \cite{sener2017active} from each individual previous site. This ignores data heterogeneity across sites and thus causes rehearsal redundancy, weakening both memorizability and generalizability.
These motivate us to integrate the schemes of both CL and DG for SMG-Learning, by exploring gradient with \emph{coordination for previous sites and invariance across various sites} for model updating, and further enforce additional inter-site diversity of representative exemplars in the replay buffer.

In this paper, we present, \emph{to our knowledge for the first}, SMG-Learning framework for continual multi-site segmentation.
Our contributions are:
1) We propose a novel Synchronous Gradient Alignment (SGA) objective with two complementary operations, including an orientational alignment that determines coordinated gradient for the previous sites in replay buffer to strengthen memorizability, and an arbitrary alignment that encourages site-invariance across randomly split subsets to promote generalizability;
2) We design a Dual-Meta algorithm which approximates the SGA objective as dual meta-objectives for simplified optimization, avoiding costly computation of the second-order derivatives required in the naive direct optimization;
3) We propose a comprehensive configuration of replay buffer, where each exemplar is sampled considering intra-site representativeness and inter-site diversity, to relieve redundancy and improve rehearsal efficiency.
We have evaluated our method on prostate MRI segmentation, using a sequential stream of multi-site datasets for model training.
Experimental results show that our method achieves synchronously high memorizability and generalizability, and clearly outperforms the state-of-the-art CL and DG methods.

\begin{figure}[t]
    \centering
    \includegraphics[width=\textwidth]{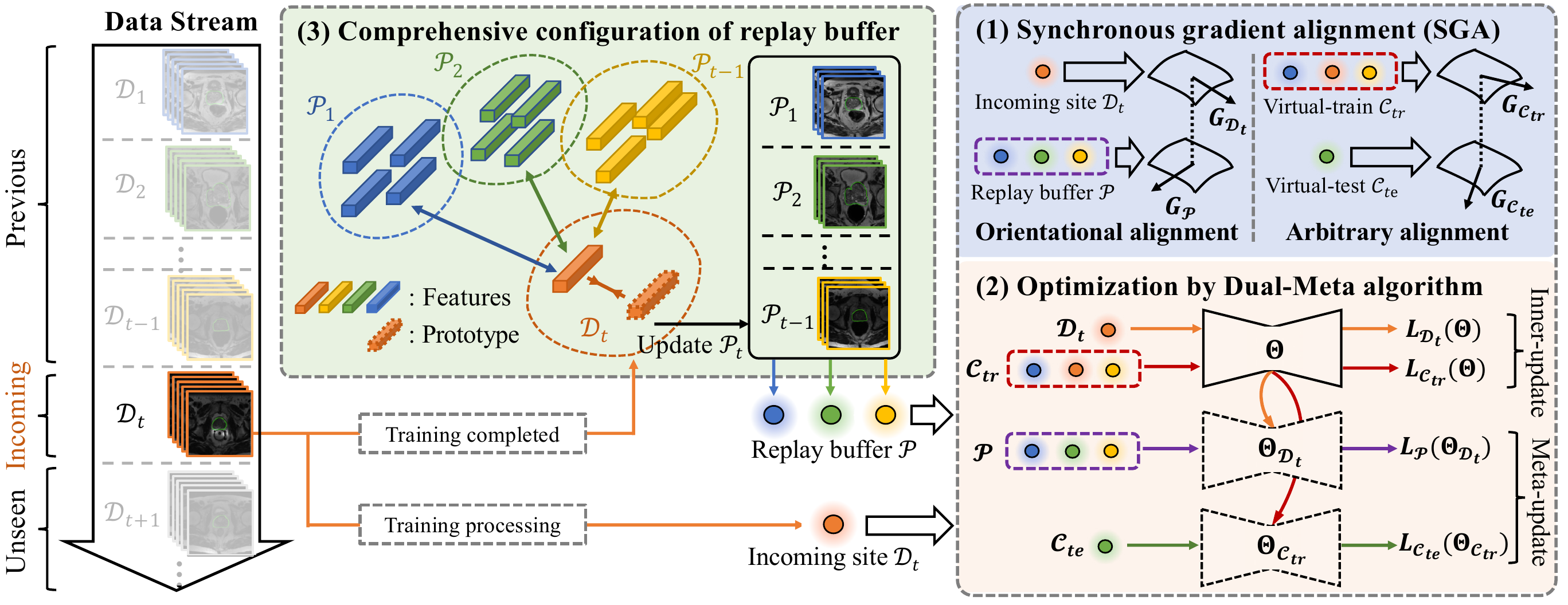} 
    \caption{Overview of our SMG-Learning framework for synchronous memorizability and generalizability. In the training process, we use a Synchronous Gradient Alignment (SGA) objective to enforce coordinated optimization for previous sites and enhance invariance across sites (Sec. \ref{sec:SGA}). This SGA objective is optimized by Dual-Meta algorithm without costly computation overhead (Sec. \ref{sec:Dual-Meta}). After completing the training process, we configure the replay buffer comprehensively for efficient rehearsal (Sec. \ref{sec:ComprehensiveSampling}).}
    \label{overview}
\end{figure}

\section{Method}
For consecutive model training, we use a sequential data stream from $T$ sites.
Specifically, in round $t \in [1, T]$ of this procedure, we can obtain only the data $\mathcal{D}_t$ of incoming site, and allow access to a small replay buffer $\mathcal{P}=\{\mathcal{P}_1,\mathcal{P}_2,...,\mathcal{P}_{t-1}\}$ that stores a small number of exemplars from previous sites for rehearsal.
Data $\{\mathcal{D}_{t+1},...,\mathcal{D}_{T}\}$ of unseen sites is not used in the training process.
Fig. \ref{overview} illustrates the proposed SMG-Learning framework,
aiming to simultaneously perform well on both previous and unseen sites during the consecutive model training process.

\subsection{Synchronous Gradient Alignment (SGA)}
\label{sec:SGA}
Learning only from data $\mathcal{D}_t$ causes overfitting on the incoming site and also cannot enable memorizability and generalizability.
Fortunately, replay buffer $\mathcal{P}$ not only allows efficient rehearsal to mitigate model forgetting, but also provides multi-site data distribution for a generalized model.
Therefore, it is crucial for SMG-Learning to collaborate the data of $\mathcal{D}_t$ and $\mathcal{P}$ for model training.

\vspace{0.5em}
\noindent
\textbf{Naive Joint Minimization (JM).}
For improved memorizability, a naive way is to adopt Joint Minimization (JM) for losses on $\mathcal{D}_t$ and $\mathcal{P}$ w.r.t. model parameters $\rm\Theta$ for maintaining performance on both incoming and previous sites.
Moreover, to strengthen generalizability, we collaborate $\mathcal{D}_t$ and $\mathcal{P}$ as a union set $\mathcal{C}$, and then randomly split it into virtual-train $\mathcal{C}_{tr}$ and virtual-test $\mathcal{C}_{te}$ subsets within each minibatch, with a naive JM defined on both subsets for the training awareness of simulated cross-site discrepancy \cite{dou2019domain}.
These JM objectives are formulated as:
\begin{equation}
L_{\text{JM}}^{\{\mathcal{D}_t,\mathcal{P}\}}({\rm\Theta})= L_{\mathcal{D}_t}({\rm\Theta}) + L_{\mathcal{P}}({\rm\Theta}), \ \ \ \text{and} \ \ \  L_{\text{JM}}^{\{\mathcal{C}_{tr},\mathcal{C}_{te}\}}({\rm\Theta})= L_{\mathcal{C}_{tr}}({\rm\Theta}) + L_{\mathcal{C}_{te}}({\rm\Theta}).
\label{eq:ERM}
\end{equation}

However, these naive JM objectives would be optimized asymmetrically,
e.g., for $L_{\text{JM}}^{\{\mathcal{D}_t,\mathcal{P}\}}({\rm\Theta})$,
a dominated decrease comes from $L_{\mathcal{D}_t}({\rm\Theta})$ which is easier to minimize without domain shift across sites \cite{li2018learning},
yet with a risk of increasing $L_{\mathcal{P}}({\rm\Theta})$.
In this case, the model learns sufficiently on the incoming site while sacrificing performance on previous sites.
In addition, the asymmetric optimization for $L_{\text{JM}}^{\{\mathcal{C}_{tr},\mathcal{C}_{te}\}}({\rm\Theta})$ fails to obtain the consistent minimization on random subsets, and thus cannot extract site-invariant features that generalize well to unseen sites.

\vspace{0.5em}
\noindent
\textbf{Definition of SGA Objective.}
To solve the problem of naive JM objectives, we need to simultaneously enforce coordinated descent of $L_{\mathcal{D}_t}({\rm\Theta})$ and $L_{\mathcal{P}}({\rm\Theta})$ and extract invariant feature representation from $L_{\mathcal{C}_{tr}}({\rm\Theta})$ and $L_{\mathcal{C}_{te}}({\rm\Theta})$.
Let us consider the gradients of losses on $\mathcal{D}_t$, $\mathcal{P}$, $\mathcal{C}_{tr}$ and $\mathcal{C}_{te}$:
\begin{equation}
G_{\mathcal{D}_t} =\frac{\partial L_{\mathcal{D}_t}({\rm\Theta})}{\partial{\rm\Theta}}, \ G_{\mathcal{P}} = \frac{\partial L_{\mathcal{P}}({\rm\Theta})}{\partial{\rm\Theta}}, \ G_{\mathcal{C}_{tr}} = \frac{\partial L_{\mathcal{C}_{tr}}({\rm\Theta})}{\partial{\rm\Theta}}, \ G_{\mathcal{C}_{te}} = \frac{\partial L_{\mathcal{C}_{te}}({\rm\Theta})}{\partial{\rm\Theta}}.
\label{eq:SGA}
\end{equation}
If $G_{\mathcal{D}_t}$ and $G_{\mathcal{P}}$ have a similar direction, i.e., inner product $G_{\mathcal{D}_t} \cdot G_{\mathcal{P}}>0$, updating model with a step along $G_{\mathcal{D}_t}$ or $G_{\mathcal{P}}$ improves performance on both $\mathcal{D}_t$ and $\mathcal{P}$, achieving coordinated optimization and thus a high memorizability for previous sites.
Moreover, if the directions of $G_{\mathcal{C}_{tr}}$ and $G_{\mathcal{C}_{te}}$ are similar, i.e., $G_{\mathcal{C}_{tr}} \cdot G_{\mathcal{C}_{te}} > 0$, the features learned by either step can be invariant to the simulated discrepancy between arbitrarily split $\mathcal{C}_{tr}$ and $\mathcal{C}_{te}$, strengthening potential generalizability.

Based on this observation, we propose to simultaneously maximize $G_{\mathcal{D}_t} \cdot G_{\mathcal{P}}$ for orientational gradient alignment between incoming site and replay buffer, and $G_{\mathcal{C}_{tr}} \cdot G_{\mathcal{C}_{te}}$ for arbitrary gradient alignment across randomly split subsets\footnote{Orientational alignment seems a special case of arbitrary alignment when the subset splitting is $\mathcal{C}_{tr}=\mathcal{D}_{t}$ and $\mathcal{C}_{te}=\mathcal{P}$, coincidently.
However, orientational alignment cannot be omitted with a risk of suffering from potential interference \cite{riemer2018learning}, as empirically shown in Appendix A, due to arbitrary alignment for other subset splittings.}.
They enforce the gradient with coordination for previous sites and invariance across various sites, benefiting to SMG-Learning.
Formally, we name our objective as Synchronous Gradient Alignment (SGA), formalized by subtracting $G_{\mathcal{D}_t}\cdot G_{\mathcal{P}}$ and $G_{\mathcal{C}_{tr}}\cdot G_{\mathcal{C}_{te}}$ from the naive JM objectives with weight $\gamma$ and $\beta$:
\begin{equation}
L_{\text{SGA}}({\rm\Theta}) = L_{\text{JM}}^{\{\mathcal{D}_t,\mathcal{P}\}}({\rm\Theta}) + L_{\text{JM}}^{\{\mathcal{C}_{tr},\mathcal{C}_{te}\}}({\rm\Theta}) -\gamma\;G_{\mathcal{D}_t} \cdot G_{\mathcal{P}} - \beta\;G_{\mathcal{C}_{tr}} \cdot G_{\mathcal{C}_{te}}.
\label{eq:SGA}
\end{equation}

\noindent
\textbf{Relationship with CL and DG methods.}
If only the orientational alignment is used, the SGA objective can be regarded as a CL variant \cite{gupta2020maml} with a difference: it ensures coordinated descent explicitly without inequality constraints \cite{lopez2017gradient}.
If we only activate the arbitrary alignment, SGA objective borrows the spirit from DG with random data splitting \cite{li2018learning} to simulate domain shift for feature invariance.
A unified combination of them in our SGA objective leverages the complementary strengths of CL and DG and thereby makes the SMG-Learning promising.

\subsection{Efficient Optimization by Dual-Meta Algorithm}
\label{sec:Dual-Meta}
A naive optimization of our SGA objective is costly since it requires to compute second-order derivatives due to the gradient inner product terms.
To solve it, we propose a Dual-Meta algorithm that approximates SGA objective as dual meta-objectives for meta-optimization to avoid expensive second-order computation.

Formally, the proposed SGA objective can be approximated as dual meta-objectives by the first-order Taylor expansion with omitted infinitesimal:
\begin{equation}
L_{\text{SGA}}({\rm\Theta}) = \underbrace{L_{\mathcal{D}_t}({\rm\Theta}) +L_{\mathcal{P}}({\rm\Theta}-\gamma G_{\mathcal{D}_t})}_{\text{the first meta-objective}} + \underbrace{L_{\mathcal{C}_{tr}}({\rm\Theta}) + L_{\mathcal{C}_{te}}({\rm\Theta}-\beta G_{\mathcal{C}_{tr}})}_{\text{the second meta-objective}}
\label{eq:dual_meta}
\end{equation}
where dual components correspond to the maximization of $G_{\mathcal{D}_t}\cdot G_{\mathcal{P}}$ and $G_{\mathcal{C}_{tr}}\cdot G_{\mathcal{C}_{te}}$ in SGA objective, respectively.
Derivation details are given in Appendix B.

Based on this formulation of dual meta-objectives \cite{li2018learning}, we perform a classical meta-optimization \cite{finn2017model} including two stages.
First, in inner-update, model parameters $\rm\Theta$ are updated on $\mathcal{D}_t$ and $\mathcal{C}_{tr}$ with step size $\gamma$ and $\beta$, respectively:
\begin{equation}
{\rm\Theta}_{\mathcal{D}_t} = {\rm\Theta}-\gamma G_{\mathcal{D}_t} \quad \text{and} \quad {\rm\Theta}_{\mathcal{C}_{tr}} = {\rm\Theta}-\beta G_{\mathcal{C}_{tr}}.
\end{equation}
Second, in meta-update, we rewrite the dual meta-objectives as $L_{\mathcal{D}_t}({\rm\Theta}) + L_{\mathcal{P}}({\rm\Theta}_{\mathcal{D}_t})$ $ + L_{\mathcal{C}_{tr}}({\rm\Theta}) + L_{\mathcal{C}_{te}}({\rm\Theta}_{\mathcal{C}_{tr}})$. Notice that loss $L_{\mathcal{P}}({\rm\Theta}_{\mathcal{D}_t})$ and $L_{\mathcal{C}_{te}}({\rm\Theta}_{\mathcal{C}_{tr}})$ are computed on $\mathcal{P}$ and $\mathcal{C}_{tr}$ with the updated parameters ${\rm\Theta}_{\mathcal{D}_t}$ and ${\rm\Theta}_{\mathcal{C}_{tr}}$, but optimized towards the original parameters $\rm\Theta$ where only the first-order derivative is required.

\subsection{Comprehensive Configuration of Replay Buffer}
\label{sec:ComprehensiveSampling}
After completing the learning round $t$, we should sample a small exemplar set $\mathcal{P}_t$ from $\mathcal{D}_t$ of incoming site, and then add it into replay buffer $\mathcal{P}$ for rehearsal in the next learning round.
For rehearsal efficacy, the selected exemplars are required with intra-site representativeness for the data distribution of $\mathcal{D}_t$.
Meanwhile, they are also demanded with inter-site diversity for exemplars of previous sites in $\mathcal{P}=\{\mathcal{P}_1,...,\mathcal{P}_{t-1} \}$ to reduce redundancy.
Therefore, we propose a comprehensive configuration of $\mathcal{P}_t$ that considers both properties for replay buffer updating.

\vspace{0.5em}
\noindent
\textbf{Intra-Site Representativeness.}
Given a subject $s_t^i \in \mathcal{D}_t$, we extract its feature vector $f_{\rm\Theta}(s_t^i)$ by average-pooling its feature map from the bottleneck layer of the model parameterized with ${\rm\Theta}$.
Its intra-site representativeness is defined as the cosine similarity, i.e., $\text{sim}(a, b)=a^{T} b /\|a\|\|b\|$, to the site prototype $\mu_t$ \cite{sener2017active}:
\begin{equation}
R(s_t^i)=\text{sim}(f_{\rm\Theta}(s_t^i),\mu_t), \qquad \text{with} \qquad \mu_t={{\sum}_{s_t^k \in \mathcal{D}_t}f_{\rm\Theta}(s_t^k)} /{N_{\mathcal{D}_t}},
\end{equation}
where $\mu_t$ is estimated as the mean feature over all $N_{\mathcal{D}_t}$ subjects in $\mathcal{D}_t$.
A high similarity $R(s_t^i)$ with respect to $\mu_t$ indicates a high representativeness for $\mathcal{D}_t$.

\vspace{0.5em}
\noindent
\textbf{Inter-Site Diversity.}
For a subject $s_t^i \in \mathcal{D}_t$ from a given individual site, inter-site diversity $V(s_t^i)$ measures its diversity w.r.t. all previous sites in replay buffer:
\begin{equation}
V(s_t^i)={\sum}_{\mathcal{P}_k\in\mathcal{P}}\text{div}(s_t^i, \mathcal{P}_k)/N_{\mathcal{P}}, \ \ \text{with} \ \ \
\text{div}(s_t^i, \mathcal{P}_k)=\underset{s_k^j\in \mathcal{P}_k}{\text{min}}\text{dsim}(f_{\rm\Theta}(s_t^i), f_{\rm\Theta}(s_k^j)),
\end{equation}
where $\text{div}(s_t^i, \mathcal{P}_k)$ denotes the diversity of $s_t^i$ w.r.t. a previous site $\mathcal{P}_k \in \mathcal{P}$ in replay buffer, and the inter-site diversity $V(s_t^i)$ is the average over all $N_{\mathcal{P}}$ previous sites.
Notably, $\text{div}(s_t^i, \mathcal{P}_k)$ is defined as the minimal feature dissimilarity, i.e., $\text{dsim}(a,b)\!=\!-\text{sim}(a,b)$, of all subjects in $\mathcal{P}_k$, which is insensitive to outliers \cite{wang2016cost}.

\vspace{0.5em}
\noindent
\textbf{Comprehensive Configuration.}
Based on $R(s_t^i)$ and $V(s_t^i)$, we adopt a hybrid measurement $H(s_t^i)$ with a weight $\lambda$ and then propose a comprehensive configuration of exemplar set $\mathcal{P}_t$ with size $N_e$ to update replay buffer.
The selected $s_t^i$ in $\mathcal{P}_t$ should have the top-$N_e$ hybrid measurement, i.e., satisfying $H(s_t^i)\geq\Gamma(N_e)$:
\begin{equation}
\mathcal{P}_t = \{s_t^i|s_t^i \in \mathcal{D}_t, H(s_t^i)\geq\Gamma(N_e)\}, \quad \text{with} \quad H(s_t^i)=R(s_t^i)+\lambda V(s_t^i).
\end{equation}

\section{Experiments}

\noindent
\textbf{Dataset.}
We employed a well-established multi-site T2-weighted MRI dataset for prostate segmentation \cite{10.1007/978-3-030-59713-9_46}: 30 subjects in RUNMC \cite{nci_isbi} (Site A), 30 subjects in BMC \cite{nci_isbi} (Site B), 19 subjects in HCRUDB \cite{LEMAITRE20158} (Site C), 13 subjects in UCL \cite{LITJENS2014359} (Site D), 12 subjects in BIDMC \cite{LITJENS2014359} (Site E) and 12 subjects in HK \cite{LITJENS2014359} (Site F).
They were acquired using different protocols with in/through plane resolution ranging from 0.25/2.2-3 to 0.625/3.6 mm.
For pre-processing, each image was resized to 384$\times$384 in axial plane and normalized to zero mean and unit variance. In each site, we used 60\%, 15\% and 25\% of subjects for training, validation and testing. This dataset is publicly released at {\url{https://liuquande.github.io/SAML/}}.

\vspace{0.5em}
\noindent
\textbf{Experimental Setting.}
For consecutive model training,
we sequentially organized the training sets of site A$\to$B$\to$C$\to$D$\to$E as a training data stream $S_{A\to E}$, while leaving site F as an unseen site without involved in training. Similarly, a reversed stream $S_{F\to B}$ with unseen site A was also validated in experiments.

\vspace{0.5em}
\noindent
\textbf{Evaluation Metrics.}
In each site, we evaluate segmentation accuracy by Dice Score Coefficient (DSC) and Average Surface Distance (ASD). Based on them, we define several specialized metrics \cite{zhang2021comprehensive,lopez2017gradient} to evaluate memorizability and generalizability.
First, for memorizability, we define a Backward Measure (BM) as the mean segmentation accuracy over previous and incoming sites for generic evaluation, and a Backward Transfer (BT) as the accuracy degradation on previous sites after learning on an incoming site for quantitative evaluation.
Second, to evaluate generalizability, we define a Forward Measure (FM) as the segmentation accuracy on unseen site, and a Forward Transfer (FT) as the gap between the accuracy on unseen site obtained before and after learning on this unseen site.
An advanced method should have high BM and FM for DSC while with low values for ASD, and its BT and FT should be as close as possible to zero.

\vspace{0.5em}
\noindent
\textbf{Implementation.}
We adopted a 2D-UNet \cite{10.1007/978-3-319-24574-4_28} as segmentation backbone due to the large variance in slice thickness among different sites.
For our SGA objective, both parameter $\gamma$ and $\beta$ were empirically set as 5$e^{-4}$ \cite{li2018learning} for suitable trade-off.
Weight $\lambda$ was set as 1.0 to comprehensively configure replay buffer, where each exemplar set contained only $N_e=2$ subjects for storage efficiency.
The network was trained by Dual-Meta algorithm, using Adam Optimizer with learning rate 5$e^{-4}$, batch size 5 and iteration number 20$K$ for inner-update and meta-update.

\vspace{0.5em}
\noindent
\textbf{Comparison with State-of-the-arts.}
We compare our SMG-Learning framework, including \textbf{SGA} (optimized by Dual-Meta algorithm) and \textbf{SGA+(c)} (integrated with comprehensive replay buffer), with the state-of-the-art methods:
1) Baselines: \textbf{FineTuning} only on the incoming site, and Joint Minimization (\textbf{JM}) using only Eq.(\ref{eq:ERM}) with representative replay buffer \cite{sener2017active};
2) CL schemes: Gradient Episodic Memory (\textbf{GEM}) \cite{lopez2017gradient} and {Continual Meta-learning (\textbf{C-Meta})} \cite{gupta2020maml} for implicitly and explicitly coordinated optimization on previous sites;
3) DG schemes: {Inter-Site Gradient Matching (\textbf{ISGM})} \cite{shi2021gradient} for feature site-invariance and {Generalized Meta-learning (\textbf{G-Meta})} \cite{li2018learning} on randomly split subsets.

\begin{table}[t]
    \caption{Comparison with state-of-the-arts for network memorizability (Mem.) and generalizability (Gen.), after training on two different data stream $S_{A \to E}$ and $S_{F \to B}$.}
    \resizebox{\textwidth}{15mm}{
    \begin{tabular}{cc|cccccccc||cccccccc}
        \hline \hline
        \multicolumn{2}{c|}{Settings} & \multicolumn{8}{c||}{Training data stream $S_{A \to E}$} & \multicolumn{8}{c}{Training data stream $S_{F \to B}$} \\ [2pt]

        \hline \hline
        \multicolumn{2}{c|}{\multirow{3}{*}{Metrics}} & \multicolumn{4}{c|}{DSC(\%)$\bm{\uparrow}$} & \multicolumn{4}{c||}{ASD(mm)$\bm{\downarrow}$} & \multicolumn{4}{c|}{DSC (\%)$\bm{\uparrow}$} & \multicolumn{4}{c}{ASD (mm)$\bm{\downarrow}$} \\

        \cline{3-18}
        \multicolumn{2}{c|}{} & \multicolumn{2}{c|}{Mem.} & \multicolumn{2}{c|}{Gen.} & \multicolumn{2}{c|}{Mem.} & \multicolumn{2}{c||}{Gen.} & \multicolumn{2}{c|}{Mem.} & \multicolumn{2}{c|}{Gen.} & \multicolumn{2}{c|}{Mem.} & \multicolumn{2}{c}{Gen.} \\

        \cline{3-18}
        \multicolumn{2}{c|}{} & BM & \multicolumn{1}{c|}{BT} & FM & \multicolumn{1}{c|}{FT} & BM & \multicolumn{1}{c|}{BT} & FM          & FT & BM & \multicolumn{1}{c|}{BT} & FM & \multicolumn{1}{c|}{FT} & BM & \multicolumn{1}{c|}{BT} & FM & FT \\

        \hline\hline
        \multicolumn{1}{c|}{\multirow{2}{*}{Baseline}} & FineTuning & 62.64 & \multicolumn{1}{c|}{-27.40} & 82.77 & \multicolumn{1}{c|}{-7.01} & 5.00 & \multicolumn{1}{c|}{3.84} & 2.40 & 1.38 & 64.20 & \multicolumn{1}{c|}{-24.07} & 67.87 & \multicolumn{1}{c|}{-24.39} & 7.57 & \multicolumn{1}{c|}{6.21} & 11.84 & 10.72 \\

        \multicolumn{1}{c|}{} & JM \cite{sener2017active} & 73.98 & \multicolumn{1}{c|}{-14.24} & 83.85 & \multicolumn{1}{c|}{-5.07} & 3.36 & \multicolumn{1}{c|}{2.01} & 2.24 & 1.14 & 73.81 & \multicolumn{1}{c|}{-13.45} & 70.39 & \multicolumn{1}{c|}{-20.67} & 3.18 & \multicolumn{1}{c|}{1.62} & 5.47 & 4.16 \\

        \hline
        \multicolumn{1}{c|}{\multirow{2}{*}{CL}} & GEM \cite{lopez2017gradient} & 79.69 & \multicolumn{1}{c|}{-9.58} & 84.82 & \multicolumn{1}{c|}{-4.47} & 2.48 & \multicolumn{1}{c|}{1.23} & 2.13 & 1.08 & 78.15 & \multicolumn{1}{c|}{-9.52} & 72.31 & \multicolumn{1}{c|}{-19.43} & 2.41 & \multicolumn{1}{c|}{0.94} & 3.96 & 2.70 \\

        \multicolumn{1}{c|}{} & C-Meta \cite{gupta2020maml} & 81.17 & \multicolumn{1}{c|}{-7.59} & 85.83 & \multicolumn{1}{c|}{-3.84} & 2.09 & \multicolumn{1}{c|}{0.81} & 1.79 & 0.72 & 80.88 & \multicolumn{1}{c|}{-6.92}  & 72.33 & \multicolumn{1}{c|}{-19.57} & 2.14 & \multicolumn{1}{c|}{0.66} & 3.68 & 2.47 \\

        \hline
        \multicolumn{1}{c|}{\multirow{2}{*}{DG}} & ISGM \cite{shi2021gradient} & 80.49 & \multicolumn{1}{c|}{-8.01} & 86.22 & \multicolumn{1}{c|}{-2.97} & 2.25 & \multicolumn{1}{c|}{0.93} & 1.54 & 0.45 & 78.60 & \multicolumn{1}{c|}{-9.03}  & 74.48 & \multicolumn{1}{c|}{-17.04} & 2.40 & \multicolumn{1}{c|}{0.97} & 3.46 & 2.23 \\

        \multicolumn{1}{c|}{} & G-Meta \cite{li2018learning} & 78.98 & \multicolumn{1}{c|}{-10.78} & 86.31 & \multicolumn{1}{c|}{-2.94} & 2.33 & \multicolumn{1}{c|}{1.17} & 1.50 & 0.43 & 77.32 & \multicolumn{1}{c|}{-10.15} & 75.06 & \multicolumn{1}{c|}{-16.70} & 2.55 & \multicolumn{1}{c|}{1.17} & 3.43 & 2.21 \\

        \hline
        \multicolumn{1}{c|}{\multirow{2}{*}{SMG-Learning}} & SGA & 81.68 & \multicolumn{1}{c|}{-7.41}  & 86.46 & \multicolumn{1}{c|}{-2.72} & 1.99 & \multicolumn{1}{c|}{0.76} & 1.42 & 0.36 & 81.72 & \multicolumn{1}{c|}{-6.17}  & 74.82 & \multicolumn{1}{c|}{-16.69} & \textbf{2.06} & \multicolumn{1}{c|}{0.64} & 3.52 & 2.30 \\

        \multicolumn{1}{c|}{} & SGA(+c) & \textbf{83.60} & \multicolumn{1}{c|}{\textbf{-5.25}} & \textbf{87.18} & \multicolumn{1}{c|}{\textbf{-2.30}} & \textbf{1.81} & \multicolumn{1}{c|}{\textbf{0.59}} & \textbf{1.34} & \textbf{0.30} & \textbf{82.35} & \multicolumn{1}{c|}{\textbf{-5.42}}  & \textbf{77.43} & \multicolumn{1}{c|}{\textbf{-14.05}} & 2.09 & \multicolumn{1}{c|}{\textbf{0.60}} & \textbf{3.20} & \textbf{1.95} \\
        \hline \hline
    \end{tabular}}
    \label{tab:comparison}
\end{table}

\begin{figure}[t]
    \includegraphics[width=\textwidth]{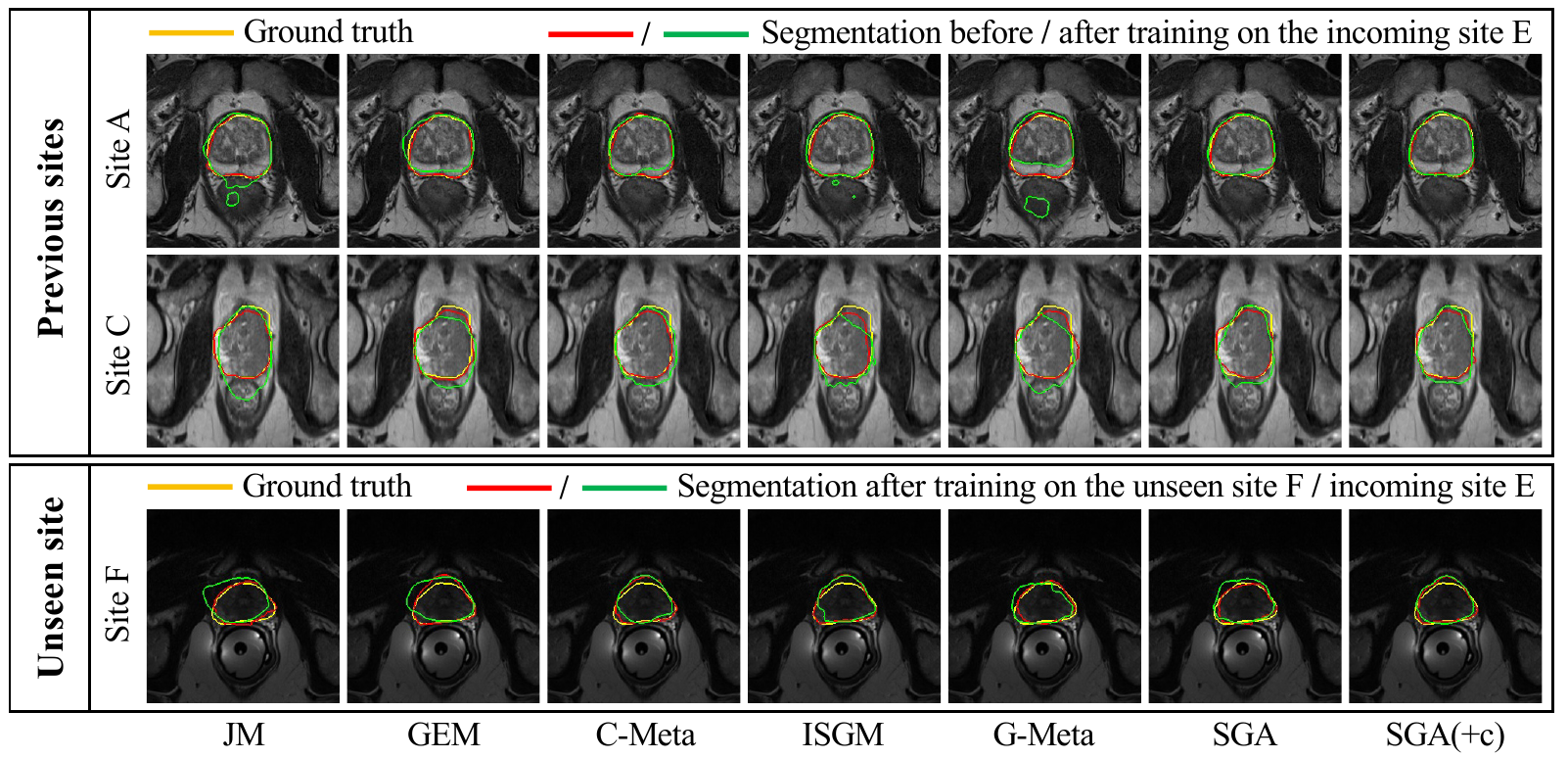}
    \caption{
    Segmentation visualization when training on stream $S_{A \to E}$.
    Upper: results on previous site A and C obtained {before (i.e., immediately after training on site A and C, respectively)} and {after training on incoming site E}; Lower: results on unseen site F obtained after training on this unseen site F and after training on the incoming site E.}
    \label{fig:seg}
\end{figure}

As listed in Table \ref{tab:comparison}, baselines show a poor memorizability with the worst BM and BT values, and a weak generalizability with the worst FM and FT values.
Although CL and DG perform better than baselines, their improvements are still limited either on memorizability or generalizability instead of in a synchronous way.
The proposed SGA consistently outperforms these methods on almost all metrics.
Moreover, SGA+(c) achieves the best results on all two experimental settings, e.g., with an obvious advantage over C-Meta by 2.34\% and 1.50\% BT for DSC and also over G-Meta by 0.64\% and 2.65\% FT, indicating both high memorizability and generalizability of our SMG-Learning framework.

Fig. \ref{fig:seg} visualizes the segmentation results on previous and unseen sites.
The red curves on different sites are the results obtained immediately after training on these individual sites.
They are close to the ground truth, revealing the high cross-site adaptation capacity of all methods.
JM degrades the performance on previous site A and C after learning on incoming site E, and causes inaccuracy on unseen site F if not involved in training.
Compared to other methods with only advantage either on previous or unseen site, our SGA achieves a synchronous superiority and especially SGA(+c) performs the best, with the highest overlap for previously obtained results and the highest accuracy on unseen site.

\begin{figure}[t]
    \includegraphics[width=\textwidth]{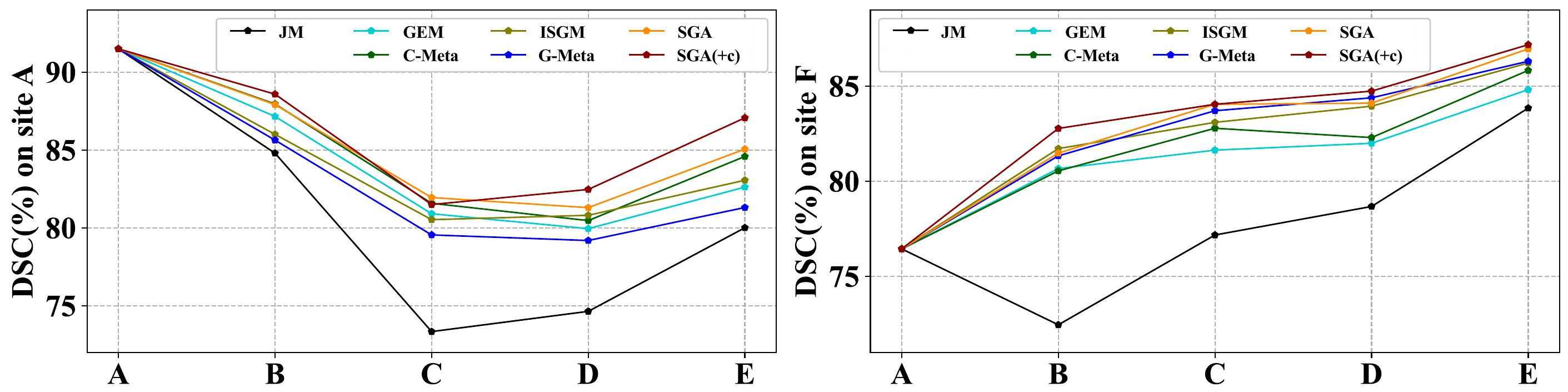}
    \caption{Curves of DSC on site A (i.e., previous site) and site F (i.e., unseen site) as more sites participate in the consecutive model training ordered by A$\to$B$\to$C$\to$D$\to$E.}
    \label{fig:lines}
\end{figure}

\vspace{0.5em}
\noindent
\textbf{Effect of Training Sequence Length.}
Fig. \ref{fig:lines} shows the DSC curves on site A and F, revealing how the network memorizability and generalizability changes, as we gradually increase the training sequence ordered by site A$\to$B$\to$C$\to$D$\to$E.
Intuitively, a longer sequence contains more complex multi-site information that would cause more optimization interference, thus aggravating model forgetting for previous site A.
Contrarily, its covered more comprehensive data distribution helps the model generalize better to unseen site F.
Notably, SGA and SGA(+c) consistently maintain higher accuracy than other methods with different lengths of training data sequence, confirming the stable efficacy of our SMG-Learning.

\vspace{0.5em}
\noindent
\textbf{Ablation Study.}
C-Meta and G-Meta can be regarded as the variants of SGA using only orientational and arbitrary alignment, respectively.
They exhibit limited advantages on either memorizability or generalizability, as shown in Table \ref{tab:comparison}.
Combining them in SGA outperforms each single component, and further using comprehensive replay buffer in SGA(+c) offers more superiority owing to its efficient rehearsal.
Moreover, SGA optimized by Dual-Meta algorithm achieves a comparable segmentation accuracy with the direct optimization while costs only 3.31\% computational time, revealing the efficiency of our Dual-Meta algorithm.

\section{Conclusion}
This paper presents a novel framework for SMG-Learning.
We propose a SGA objective, cooperating with a comprehensive configuration of replay buffer, to enforce coordinated optimization for previous sites and feature invariance across sites.
It is optimized by Dual-Meta algorithm without costly second-order computation.
In the future, it is of interest to extend our method to federated learning with decentralized data from different sites for better privacy protection.

\vspace{1.0em}
\noindent
\textbf{Acknowledge} This work was supported in part by National Natural Science Foundation of China (grant number 62131015), and Science and Technology Commission of Shanghai Municipality (STCSM) (grant number 21010502600).

\bibliographystyle{splncs04}
\bibliography{ref.bib}

\end{document}